# Supporting Design Decisions in Rule-Based Model Transformations

Dejan Stojimirović, University of Belgrade, Faculty of Organizational Sciences, Belgrade, Serbia, dejan.stojimirovic@fon.bg.ac.rs, 0000-0003-0444-295X,

Siniša Nešković, IT University of Copenhagen, Copenhagen, Denmark, sinn@itu.dk, 0000-0002-8309-5847

**Abstract**

Model Driven Engineering relies on model transformations to automate the derivation of target models or source code from source models. However, the design decisions that govern how source elements are mapped to target artifacts typically remain embedded in the transformation source code, limiting flexibility, reusability, and traceability.

This paper proposes an approach to explicitly model and manage design decisions in rule-based model transformations. Design decisions are separated from transformation implementation through three mechanisms. First, a decision model captures design decisions and their options independently of any source modeling language for a given transformation domain. Second, a binding model connects these decisions to specific metamodel concepts in a given source modeling language, enabling reuse of the same design knowledge across transformations from similar languages. Third, a configuration model records the specific option chosen for each applicable source model element, with defaults pre-selected automatically. During execution, variability points in transformation rules are resolved dynamically according to configured choices. A trace model records which rules and options were applied to produce each target element.

We establish a formal mathematical framework that defines the core concepts of variability-based transformations and proves key properties of the resulting transformations. The formal concepts are realized in a practical architecture of four interconnected artifact models. We demonstrate practical feasibility by extending an existing embedded domain-specific language for model-driven engineering with variability support and illustrate the approach with a complete ER-to-Relational transformation example.

# 1. Introduction

Model Driven Engineering (MDE) promotes models as the primary artifacts of software development and relies on model transformations to systematically derive target models or source code from higher-level abstractions [1], [2]. Although models are meant to capture system knowledge, model transformations themselves remain an exception to this principle. Typically written in specialized transformation languages as textual source code, the mapping logic they encode exists as code rather than as explicit models, contradicting MDE's core philosophy.

Most transformations involve design choices. A single modeling construct can legitimately be transformed in multiple ways, depending on architectural patterns, target platform constraints, or domain-specific requirements. Yet traditional transformation approaches hardcode one particular solution, embedding these decisions in the transformation logic with no clear separation between core mapping logic and context-specific choices. Without explicit documentation of why specific mapping strategies were chosen, developers have difficulty understanding transformation behavior or adapting it to new contexts. Transformations cannot easily be reused across projects, their design rationale remains implicit, and modifying them might lead to errors that are hard to find.

To address these limitations, we propose a systematic approach to explicitly model and manage design decisions in model transformations. Instead of being embedded in transformation code, design decisions are treated as first-class modeling artifacts. Our approach separates design knowledge from transformation implementation through three mechanisms: (1) A decision model captures relevant design decisions for a transformation domain independently of any source modeling language, enabling reuse of design knowledge across transformations from different languages within the same domain. (2) A binding model connects design decisions to specific metamodel concepts in a given source modeling language. (3) A configuration model records the chosen option for each applicable source model element, with defaults pre-selected automatically. This allows different elements of the same type to be transformed using different strategies. During execution, the transformation engine dynamically selects rule variants based on configured choices, and a trace model captures which source elements were transformed to which target elements, which rules were applied, and which options guided each step.

Our approach is not restricted to a specific transformation language. Instead, it requires that rule-based languages support variability points with selectable alternatives in their transformation rules. To establish these requirements, we present a formal mathematical framework that defines the core concepts of variability-based transformations and shows their key properties: well-definedness, separation, conservative extension, and guaranteed validity of default configurations.

The paper makes four contributions:(1) a formal mathematical framework that precisely defines the core concepts and proves key properties of variability-based transformations; (2) a practical architecture of four artifact models that realizes the formal framework for managing design decisions in model transformations; (3) a validation through an implementation that extends Synthesis, an embedded DSL for model-driven engineering described in [3], with variability support; and (4) a complete ER-to-Relational illustrative example that demonstrates the full lifecycle of the approach.

The remainder of the paper is organized as follows. Section 2 provides background on design decisions and model transformations, and presents a motivating example. Section 3 establishes the formal mathematical framework. Section 4 realizes the formal concepts into the four-model architecture. Section 5 describes the realization in the Synthesis embedded DSL. Section 6 demonstrates the approach through a complete illustrative example. Section 7 discusses related work, and Section 8 concludes with limitations and future directions.

# 2. Background and Motivating Example

## 2.1 Design Decisions and Variability Modeling

Software development involves design decisions at every abstraction level, from high-level architectural choices to detailed implementation strategies. These decisions influence system properties and constrain future evolution, yet they often remain implicit in the resulting artifacts [4]. Jansen and Bosch [5] refer to this loss of design rationale as "knowledge vaporization".

Various techniques have been proposed to address this problem, from explicit decision documentation (Architecture Decision Records, decision views [6], [7]) to variability modeling approaches rooted in software product line engineering [8], [9]. In variability modeling, design decisions are represented as variation points, each offering a set of alternatives. These models can be embedded into design artifacts using annotations (e.g., JPA annotations for object-relational mapping [10]) or maintained as separate, reusable models [11], [12].

Model transformations are specified by dedicated transformation languages [13], e.g., declarative approaches such as QVT Relations [14] and Triple Graph Grammars [15] or imperative languages such as QVT Operational [14] and ATL [16]. Regardless of the paradigm, these languages generally embed design choices directly into transformation code, with little support for expressing alternative mappings or documenting the rationale behind chosen strategies [17]. While variability modeling has received considerable attention in software product lines, less attention has been given to the design decisions embedded in transformations. Existing approaches [37], [38], [39] address aspects of transformation variability but do not treat design decisions as independent, first-class artifacts that can be defined separately from both the source modeling language and the transformation code.

## 2.2. Motivating Example

To show an example of design decisions that arise in transformations, we use a common scenario: mapping an entity-relationship model to a relational schema. The source model describes entities, attributes, and relationships. The target model consists of tables, columns, and foreign key constraints. While the transformation goal is clear, the mapping process involves several non-trivial design choices.

Figure 1 provides an example where Person and Company entities are linked through an Employment relationship with roles WorksIn (0..1) and Hires (0..M).

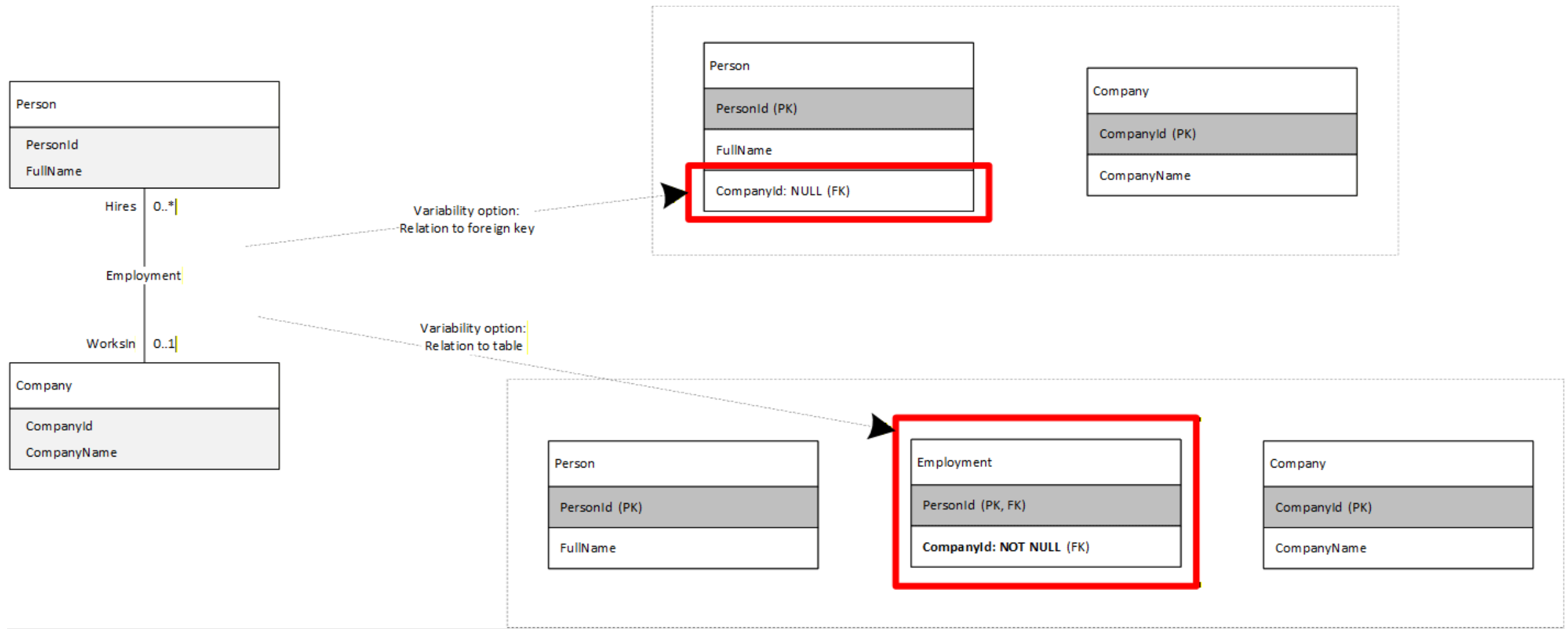

*Figure 1 Design alternatives for 0..1 to many relationship mapping*

This relationship can be mapped to a relational schema in at least two ways: (1) adding a foreign key column in the Person table referencing Company; or (2) creating a separate WorksIn table with foreign keys to both Person and Company.

Each approach has different tradeoffs. The foreign key solution simplifies the schema but introduces null values when a person is not employed. The join table approach avoids nulls and is more consistent with normalization principles, but increases schema complexity.

Similarly, other decisions arise throughout the transformation. Even seemingly simple choices, such as naming conventions for tables and columns, reflect design decisions that remain undocumented and inflexible in typical transformation implementations.

# 3. Formal Framework for Variability-Based Transformations

This section establishes the formal foundations of our approach through a mathematical framework that defines the concepts of variability-based model transformations and shows key properties that any conforming implementation must satisfy. The framework has two purposes: (1) it provides rigorous definitions of the concepts and their interrelationships, and (2) it specifies the structural requirements that any rule-based transformation language must satisfy to support variability-based transformations.

## 3.1. Design Decision Space

A design decision space captures the complete set of decisions relevant to a given transformation domain, independently of any specific source modeling language.

**Definition 1 (Design Decision Space).** A *design decision space* is a tuple $\boldsymbol{D} = (I, O, \delta, def, scope)$ where:

- $I$ is a finite, non-empty set of **issues** (design decisions),
- $O$ is a finite set of **options**,
- $\delta : I \rightarrow \wp(O)$ maps each issue to its set of available options, with $|\delta(i)| \geq 2$ for all $i \in I$,
- $def : I \rightarrow O$ assigns a default option to each issue, with $def(i) \in \delta(i)$ for all $i \in I$,
- $scope : I \rightarrow$ {model, element} classifies each issue as either model-level or element-level.
- We write $I_M = \{i \in I \mid scope(i) = \text{model}\}$ for the set of model-level issues and $I_E = \{i \in I \mid scope(i) = \text{element}\}$ for element-level issues. By definition, $I = I_M \cup I_E$ and $I_M \cap I_E = \emptyset$. Furthermore, the option sets are pairwise disjoint: for all $i_1, i_2 \in I$ with $i_1 \neq i_2$, $\delta(i_1) \cap \delta(i_2) = \emptyset$. This ensures that each option *uniquely* identifies its originating issue.

Model-level issues affect the entire transformation globally (e.g., naming conventions), while element-level issues are resolved independently for each applicable source model element (e.g., how to map a specific relationship). The requirement $|\delta(i)| \geq 2$ ensures that every issue represents a genuine design decision with at least two alternatives.

## 3.2. Metamodels and Models

**Definition 2 (Metamodel).** A metamodel is a tuple $\Sigma$ = (MC, props) where:

- MC is a finite, non-empty set of metaclasses that define the types of elements available in the modeling language,
- props : MC $\rightarrow \wp$(Prop) assigns to each metaclass its set of properties (attributes, references, and structural features).

We use $\Sigma$ to denote a metamodel generically. When distinguishing between the metamodels involved in a transformation, we write Σ_S for the source metamodel and Σ_T for the target metamodel.

**Definition 3 (Model).** A model conforming to metamodel $\Sigma$ = (MC, props) is a tuple M = (E, $\tau$, val) where:

- E is a finite set of model elements,
- $\tau$ : E $\rightarrow$ MC is a typing function that assigns each element its metaclass, establishing the conformance relationship between M1 elements and M2 metaclasses,
- val assigns property values to elements, such that for each e $\in$ E and each property p $\in$ props($\tau$(e)), val(e, p) yields the value of property p for element e.

## 3.3. Binding Model

The design decision space (Definition 1) is deliberately independent of any specific modeling language, while metamodels and models (Definitions 2–3) are independent of design decisions. A separate mechanism is needed to connect them: to specify which design decisions are relevant when transforming instances of specific metaclasses, and under what conditions. This connection is established through a binding model, which operates as a cross-level bridge, defined in terms of metaclasses at the M2 level, but with applicability conditions evaluated on model elements at the M1 level. By maintaining bindings as a separate artifact, the same design decision space can be reused across different modeling languages by creating appropriate binding models for each.

**Definition 4 (Binding).** A binding model for a design decision space $D = (I, O, \delta, def, scope)$ and a source metamodel $\Sigma_S = (MC, props)$ is a finite set $B = \{b_1, \ldots, b_n\}$ where each binding $b_k = (\sigma_k, I_k, \varphi_k)$ consists of:

- $\sigma_k \in MC$, a metaclass from the source metamodel (M2 level) to which the binding applies,
- $I_k \subseteq I$, a non-empty set of design decisions from D that are relevant when transforming instances of this metaclass,
- $\varphi_k : \{e \in E \mid \tau(e) = \sigma_k\} \rightarrow \{true, false\}$, a condition predicate evaluated on M1 model elements that are instances of σ_k. The predicate may reference property values of the element via val to determine whether the binding applies to a specific instance.

For example, a binding might connect the "relationship mapping" decision to the Relationship metaclass in the ER metamodel, with a condition predicate that checks whether a specific relationship instance has 0..1 to 0..M multiplicity. The same decision could be bound, through a separate binding model, to the Association metaclass in a UML metamodel.

We define the function applicable : $E \rightarrow \wp(I)$ that determines which design decisions apply to a given model element:

$$applicable(e) = \bigcup \{ I_k \mid b_k \in B, \sigma_k = \tau(e), \varphi_k(e) = true \}$$

## 3.4. Configuration

A configuration captures the actual choices made for a specific transformation task. It bridges the abstract decision space with a concrete source model, recording which option is selected for each applicable design decision.

**Definition 5 (Configuration).** A *configuration* for source model $\boldsymbol{M} = (E, \tau, val)$ conforming to $\Sigma_S$, given design decision space $\boldsymbol{D}$ and binding $\boldsymbol{B}$, is a pair $\boldsymbol{C} = (C_G, C_E)$ where:

- $C_G : I_M \rightarrow O$ assigns a **global choice** for each model-level issue, with $C_G(i) \in \delta(i)$,

– $C_E : E \times I_E \rightharpoonup O$ is a partial function assigning an **element-specific choice** for each applicable element-level issue, with $C_E(e, i) \in \delta(i)$ wherever defined.

To provide a uniform interface for querying choices regardless of issue level, we define a unified **choice function** $choice : E \times I \rightharpoonup O$ as:

- $choice(e, i) = C_G(i)$ if $i \in I_M$
- $choice(e, i) = C_E(e, i)$ if $i \in I_E$ and $(e, i) \in \mathrm{dom}(C_E)$

Note that for model-level issues, the choice is independent of the element *e*, reflecting the global nature of these decisions.

**Definition 6 (Valid Configuration).** A configuration $\boldsymbol{C} = (C_G, C_E)$ is *valid* with respect to $\boldsymbol{M}$, $\boldsymbol{D}$, and $\boldsymbol{B}$ if and only if:

- (Completeness of global choices) For every $i \in I_M$: $C_G(i)$ is defined and $C_G(i) \in \delta(i)$.
- (Completeness of element choices) For every $e \in E$ and every $i \in applicable(e) \cap I_E$: $C_E(e, i)$ is defined and $C_E(e, i) \in \delta(i)$.
- (No extraneous choices) $C_E(e, i)$ is defined only when $i \in applicable(e) \cap I_E$.

## 3.5. Transformation Specification

A variability-based transformation specification extends the standard notion of rule-based transformations by introducing variability points and variability options alongside regular rules.

**Definition 7 (Transformation Specification).** A *variability-based transformation specification* is a tuple $\boldsymbol{T} = (R, V, \Omega, assoc, opts, P, inv, r_0, \Sigma_S, \Sigma_T)$ where:

- $R$ is a finite set of **regular rules**, each implementing deterministic transformation logic,
- $V$ is a finite set of **variability points**, each representing a point where a design decision must be resolved,
- $\Omega$ is a finite set of **variability options**, each implementing a concrete transformation strategy,
- $assoc : V \rightarrow I$ maps each variability point to its associated design decision from $\boldsymbol{D}$,
- $opts : V \rightarrow \wp(\Omega)$ maps each variability point to its set of option rules, with $|opts(v)| \geq 2$ for all $v \in V$,
- $P : \Omega \rightarrow Pred$ assigns a predicate to each variability option (see Definition 9),
- $inv \subseteq (R \cup \Omega) \times (R \cup V)$ is the **invocation relation**, specifying which rules may invoke which other rules,
- $r_0 \in R$ is the designated **top-level rule** that initiates the transformation,
- $\Sigma_S$ and $\Sigma_T$ are the **source and target metamodels** (Definition 2) to which the source and target models must conform.

The sets $R$, $V$, and $\Omega$ are pairwise disjoint. Furthermore, the option sets are pairwise disjoint: for all $v_1$, $v_2 \in V$ with $v_1 \neq v_2$, $opts(v_1) \cap opts(v_2) = \emptyset$, ensuring each variability option belongs to exactly one variability point.

The invocation relation inv is constrained according to the following architectural principle.

**Definition 8 (Rule Classification and Invocation Constraint).** The rules of a transformation specification are classified into two overlapping categories:

- ***Executable*** $= R \cup \Omega$ — rules that contain transformation logic and can be executed,
- ***Composable*** $= R \cup V$ — rules that can be invoked by other rules.

The **invocation constraint** requires that executable rules may only invoke composable rules:

$$\text{inv} \subseteq \text{Executable} \times \text{Composable} = (\text{R} \cup \Omega) \times (\text{R} \cup \text{V})$$

Since $\Omega \cap Composable = \emptyset$ (variability options are executable but not composable), no rule can directly invoke a variability option. Variability options can only be reached through the resolution mechanism at their parent variability point. This ensures that all design decision resolution is mediated by the transformation engine.

**Definition 9 (Predicate Language).** The set *Pred* of predicates is defined inductively:

- (Atomic) Option($o$) for $o \in O$ is a predicate,
- (Default) Default() is a predicate,
- (Negation) if $p \in Pred$, then $\neg p \in Pred$,
- (Conjunction) if $p_1, p_2 \in Pred$, then $p_1 \wedge p_2 \in Pred$,
- (Disjunction) if $p_1, p_2 \in Pred$, then $p_1 \vee p_2 \in Pred$.

Predicates serve as guards on variability options. The atomic predicate Option($o$) references a specific option from the design decision space. The Default() predicate serves as a fallback when no other predicate evaluates to true.

## 3.6. Execution Semantics

The central mechanism of execution is the evaluation of predicates against a configuration to resolve variability points.

**Definition 10 (Predicate Evaluation).** Given a valid configuration $\boldsymbol{C}$ and a source element $e \in E$, the evaluation function $eval : Pred \times \boldsymbol{C} \times E \rightarrow \{\text{true, false}\}$ is defined recursively:

- $eval(\text{Option}(o), \boldsymbol{C}, e) = \text{true}$ iff $\exists\, i \in I$ such that $o \in \delta(i)$ and $choice(e, i) = o$
- $eval(\neg p, \boldsymbol{C}, e) = \neg eval(p, \boldsymbol{C}, e)$
- $eval(p_1 \wedge p_2, \boldsymbol{C}, e) = eval(p_1, \boldsymbol{C}, e) \wedge eval(p_2, \boldsymbol{C}, e)$
- $eval(p_1 \vee p_2, \boldsymbol{C}, e) = eval(p_1, \boldsymbol{C}, e) \vee eval(p_2, \boldsymbol{C}, e)$

The evaluation of Option(o) relies on the fact that option sets are pairwise disjoint (Definition 1): since each option $o$ belongs to exactly one issue $i$, the existential quantifier is uniquely determined. The evaluation simply checks whether the configuration's choice for the relevant issue matches the referenced option.

**Definition 11** (Resolution Function). Given a variability point $v \in V$, a valid configuration $\boldsymbol{C}$, and a source element $e \in E$, the **resolution set** is:

$resolve(v, \boldsymbol{C}, e) = \{ \omega \in opts(v) \mid P(\omega) \neq \text{Default}() \wedge eval(P(\omega), \boldsymbol{C}, e) = \text{true} \}$

If $resolve(v, \boldsymbol{C}, e) = \emptyset$ and there exists $\omega_d \in opts(v)$ with $P(\omega_d) = \text{Default}()$, then the default option is selected: $resolve(v, \boldsymbol{C}, e)$ is redefined to $\{\omega_d\}$. The Default() predicate thus serves as a fallback mechanism, activated only when no explicitly matched option is found.

Definition 12 (Execution Trace). An *execution trace* for transformation $\boldsymbol{T}$ on source model $\boldsymbol{M}$ with configuration $\boldsymbol{C}$ is a sequence $Tr = \langle tr_1, \ldots, tr_n \rangle$ where each trace entry $tr_k = (e_k, r_k, t_k, \omega_k, c_k)$ records:

- $e_k \in E$: the source element being transformed,
- $r_k \in R \cup V$: the invoked rule (regular rule or variability point),
- $t_k$: the produced target element(s),
- $\omega_k \in \Omega \cup \{\perp\}$: the selected variability option ($\perp$ for regular rules),
- $c_k$: the choice from $\boldsymbol{C}$ that guided the selection ($\perp$ for regular rules).

When r_k ∈ V (a variability point), the trace tuple also records the resolved option ω_k and the configuration choice c_k = choice(e_k, assoc(r_k)) that determined the resolution.

## 3.7. Properties

Here we define key properties of variability-based transformation specifications, which characterize the conditions under which transformations produce unambiguous results.

**Theorem 1(Well-definedness).** A variability-based transformation specification T is well-defined under configuration C if, for every variability point v ∈ V and every applicable source element e ∈ E:

$$|\text{resolve}(v, C, e)| = 1$$

That is, exactly one variability option is selected at each variability point.

*Sufficient conditions*:

(a) **Determinism**. If for every v ∈ V, the non-default predicates of its options are pairwise mutually exclusive, that is, for any ω_1 ≠ ω_2 ∈ opts(v) with P(ω_1) ≠ Default() and P(ω_2) ≠ Default(), there exists no valid configuration C and element e for which eval(P(ω_1), C, e) = eval(P(ω_2), C, e) = true, then |resolve(v, C, e)| ≤ 1 for all valid configurations.

(b) **Completeness**. If every variability point $v \in V$ has a default option $\omega_d \in opts(v)$ with $P(\omega_d) = Default()$, then $|resolve(v, C, e)| \geq 1$ for all valid configurations.

Proof.

(a) Mutual exclusivity of predicates directly implies that at most one non-default predicate evaluates to true for any given configuration and element. Since resolve(v, C, e) selects only options whose predicates evaluate to true (excluding default options from the initial resolution), $|resolve(v, C, e)| \leq 1$.

(b) By Definition 11, if no non-default predicate evaluates to true, the default option is selected. Therefore $resolve(v, C, e) \neq \emptyset$ for every variability point and element. Combining (a) and (b): $|resolve| \leq 1$ and $|resolve| \geq 1$ together imply $|resolve| = 1$. □

In practice, variability options typically reference individual design options of their associated decision. Under this natural assumption, mutual exclusivity follows directly from the definition of valid configurations, as shown next.

Corollary 1. When each variability option's predicate is a single atomic Option(o) reference, and the associated design decision requires exactly one selected option per issue (which is guaranteed by the definition of a valid configuration), then the predicates are automatically mutually exclusive, and the transformation is well-defined under all valid configurations.

Proof. Let $v \in V$ with $assoc(v) = i$. For any two distinct options $\omega_1, \omega_2 \in opts(v)$ with predicates $Option(o_1)$ and $Option(o_2)$, where $o_1 \neq o_2$. Since each $\omega \in opts(v)$ references an option in $\delta(assoc(v))$, both $o_1, o_2 \in \delta(i)$. A valid configuration selects exactly one option for issue i, so choice(e, i) cannot simultaneously equal both $o_1$ and $o_2$. Therefore, at most one of these predicates evaluates to true. By Theorem 1, the transformation is well-defined. □

This corollary is practically significant because it covers the common case in transformation design, as illustrated in our running example (Section 6), where each variability option simply checks for one option of one design decision.

Proposition 1 (Separation). The invocation constraint (Definition 8) ensures that for any execution trace $Tr = \langle tr_1, \ldots, tr_n \rangle$, every trace tuple tr_k where $\omega_k \neq \bot$ (i.e., a variability option was executed) corresponds to an invocation of a variability point v with $r_k \in V$. All variability resolution is mediated by the configuration.

Proof. By the invocation constraint, executable rules may only invoke composable rules. Since $\Omega \cap Composable = \emptyset$ (variability options are executable but not composable), no rule can directly invoke a variability option. Therefore, variability options can only be reached through the resolution mechanism at their parent variability point, which necessarily consults the configuration. □

This property ensures that all design decision resolutions are determined exclusively by the configuration. A transformation rule cannot bypass the variability mechanism by directly invoking an option.

Theorem 2 (Conservative Extension). Every standard (non-variable) transformation specification is a degenerate case of a variability-based specification. Conversely, a well-defined variability-based specification can be reduced to a standard specification under any valid configuration.

Formally: (a) A standard transformation specification $T_{std} = (R, r_0, \Sigma_S, \Sigma_T)$ corresponds to the variability-based specification $T = (R, \emptyset, \emptyset, \emptyset, \emptyset, \emptyset, inv, r_0, \Sigma_S, \Sigma_T)$ where $V = \Omega = \emptyset$. (b) Given a valid configuration C for which T is well-defined, the specification can be reduced to a standard specification $T_C$ by replacing each variability point v with the unique $\omega \in resolve(v, C, e)$ for each applicable element e.

Proof. (a) When $V = \Omega = \emptyset$, the specification contains only regular rules and no variability points, so the resolution mechanism is never invoked. The transformation executes exactly as a standard rule-based transformation. (b) By Theorem 1 (well-definedness), $|resolve(v, C, e)| = 1$ for every variability point and element. For each invocation of a variability point on a specific source element, the unique resolved option replaces that invocation. The resulting specification contains only executable rules with deterministic behavior and is structurally equivalent to a standard transformation specification. □

Proposition 2 (Default Configuration Validity). Given a source model M conforming to $\Sigma_S$, design decision space D, and binding B, the default configuration $C_{def}$ is defined by:

– $C_G(i) = def(i)$ for all $i \in I_M$

– $C_E(e, i) = def(i)$ for all $e \in E$ and $i \in applicable(e) \cap I_E$

is a valid configuration.

Proof. By construction, every global choice assigns the default option $def(i) \in \delta(i)$, satisfying the global completeness requirement. For every element e and applicable issue $i \in applicable(e) \cap I_E$, the choice $C_E(e, i) = def(i)$ is defined and $def(i) \in \delta(i)$, satisfying element completeness. No extraneous choices are created, as choices are generated only for applicable issues. Therefore, $C_{def}$ satisfies all three conditions of Definition 6. □

This proposition guarantees that the automatic configuration generation process (described in Section 4) always produces a valid configuration.

## 3.8. Discussion

**Relationship between components**. The definitions show a dependency structure among the formal concepts: the design decision space D is independent, bindings B depend on D and the source metamodel $\Sigma_S$, configurations C depend on B and the source model M, and execution traces depend on all of the above and the transformation specification T.

**Representation agnosticism**. The framework is deliberately agnostic about how its concepts are encoded. A model element $e \in E$ could be represented as an object in an object-oriented language, a node in a graph database, or a function in a functional framework. What matters is that the semantic requirements captured by the definitions are satisfied, making the framework applicable across different implementation paradigms.

**Practical implications**. The formal results have practical consequences for transformation engineers. Corollary 1 shows that the most common predicate pattern (one atomic Option() reference per variability option) automatically guarantees well-definedness. Theorem 1 also provides a straightforward recipe for ensuring completeness: include a default option for every variability point. Proposition 1 ensures that all design decisions are resolved exclusively through the configuration. Theorem 2 shows that adopting the framework does not disrupt existing non-variable transformations. Proposition 2 guarantees that auto-generated default configurations are always valid.

**Conformance requirements**. For a rule-based transformation language and engine to support the framework, it must provide concrete syntax mechanisms to express the concepts defined here. At a minimum, the language must support:

(i) marking rules as variability points or variability options (e.g., through decorators, annotations, or dedicated constructs),
(ii) associating predicates with variability options that reference design decision options, and
(iii) a rule dispatch mechanism that evaluates predicates against a provided configuration to resolve variability points.

# 4. Model Architecture

This section realizes the formal framework of Section 3 as a practical architecture of four interconnected models described through UML class diagrams.

Figure 2 depicts these models and their dependencies across the transformation lifecycle, which consists of design time and application (execution) time. Two human roles participate in the lifecycle. Transformation engineers (TE) create reusable design time artifacts, while application developers (AD) review and use artifacts that are auto-generated during application time. The term "application time" refers to the stage when the transformation is applied to a specific source model. Although the Configuration model is generated at this stage, reviewing and modifying configuration choices is itself a design activity. Table 1 summarizes each model's formal basis, purpose, lifecycle stage, and responsible role.

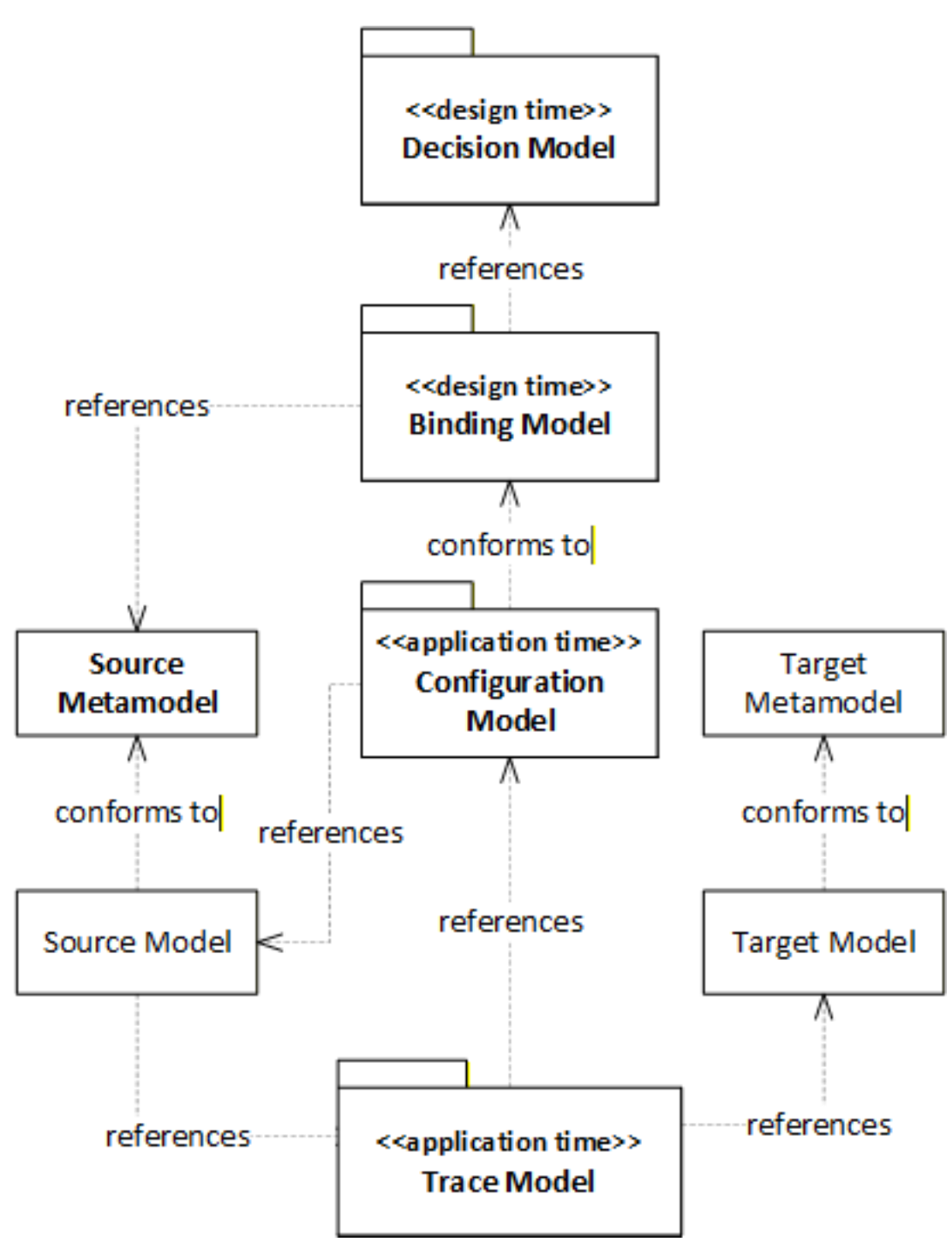

*Figure 2 Model Architecture*

| Model | Formal Basis | Purpose | Stage | Role |
|---|---|---|---|---|
| Decision | Definition 1 | Captures design issues, options, defaults, and scope (model-level vs. element-level) | Design time | TE creates |
| Binding | Definition 4 | Maps design issues to metamodel concepts (M2) | Design time | TE creates |
| Configuration | Definitions 5, 6 | Records actual choices for a specific source model | Application time | Auto-gen., AD reviews |
| Trace | Definition 8,12 | Records execution tuples (source, rule, target) with references to configuration choices | Application time | Auto-gen. |

*Table 1 Overview of the four architectural models*

The following subsections detail each model. In the class diagrams (Figures 3 through 6), each model includes a root class serving as a named container for its elements.

## 4.1. Decision model

The Decision model provides a concrete structural representation of the design decision space D = (I, O, opt, def, scope) from Definition 1. Figure 3 presents the UML class diagram.

The root DecisionModel class is a named container for all design decisions within a transformation domain. The central class is Issue, corresponding to an element $i \in I$. Each Issue has a *name*, *description*, and a *scope* enumeration (MODEL_LEVEL or ELEMENT_LEVEL) realizing the scope function from Definition 1. Model-level issues affect the entire transformation globally (e.g., naming conventions), while element-level issues are resolved per applicable source element (e.g., relationship implementation strategy).

Each Issue aggregates two or more Option instances, realizing $\delta: I \rightarrow \wp(O)$ with the constraint $|\delta(i)| \geq 2$. Each Option has a *name*, *description*, *isDefault* flag, and optional *pros* and *cons*. Exactly one Option per Issue has isDefault = true, realizing $def(i) \in \delta(i)$.

Each Option is associated with an OptionType, providing a typed identifier for referencing in predicates (Definition 9). The pros and cons attributes support design rationale documentation beyond the formal framework.

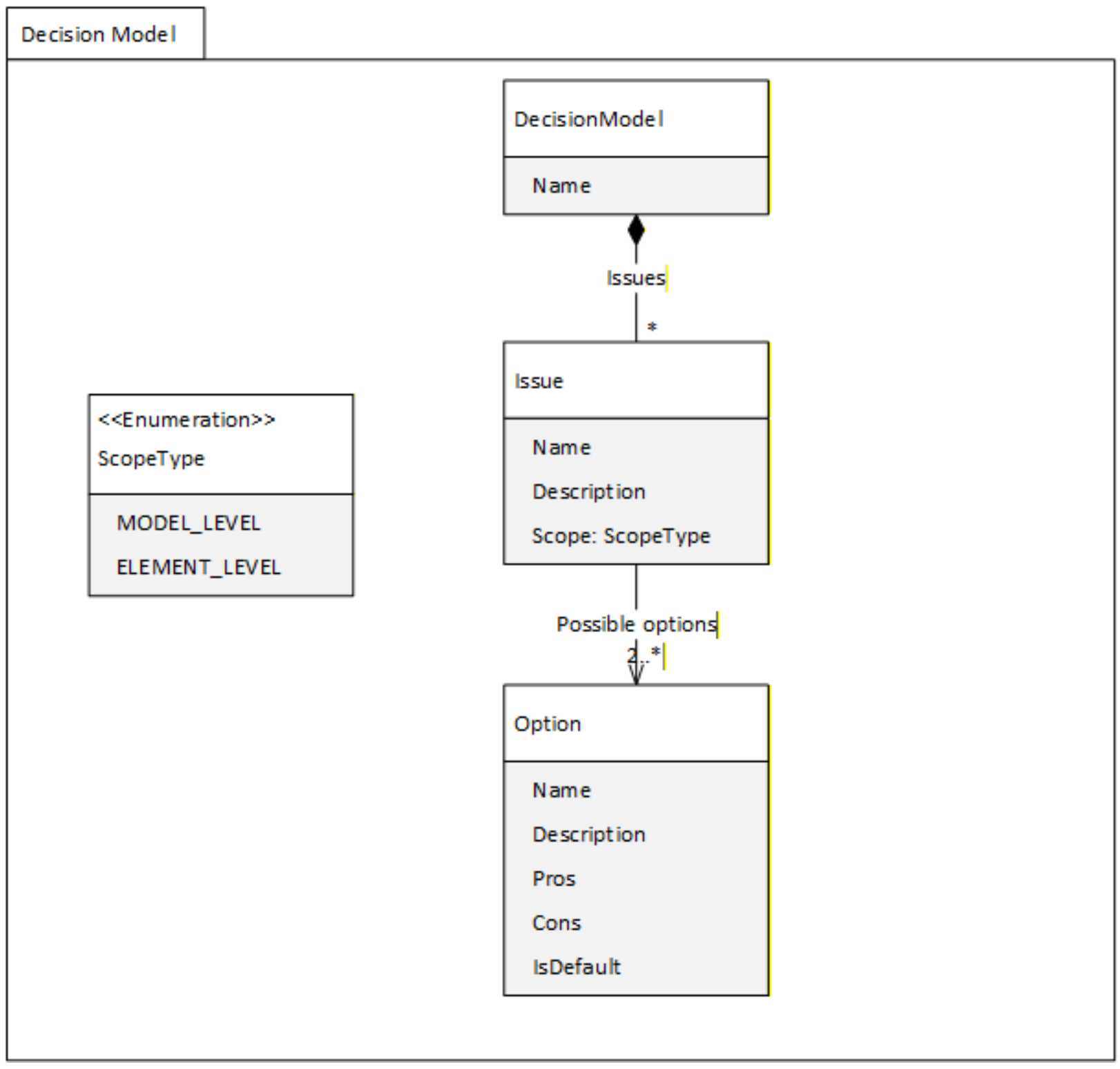

*Figure 3 Decision Model*

## 4.2. Binding Model

The Binding Model connects design decisions to specific concepts in a source modeling language, realizing Definition 4. Figure 4 presents the UML class diagram.

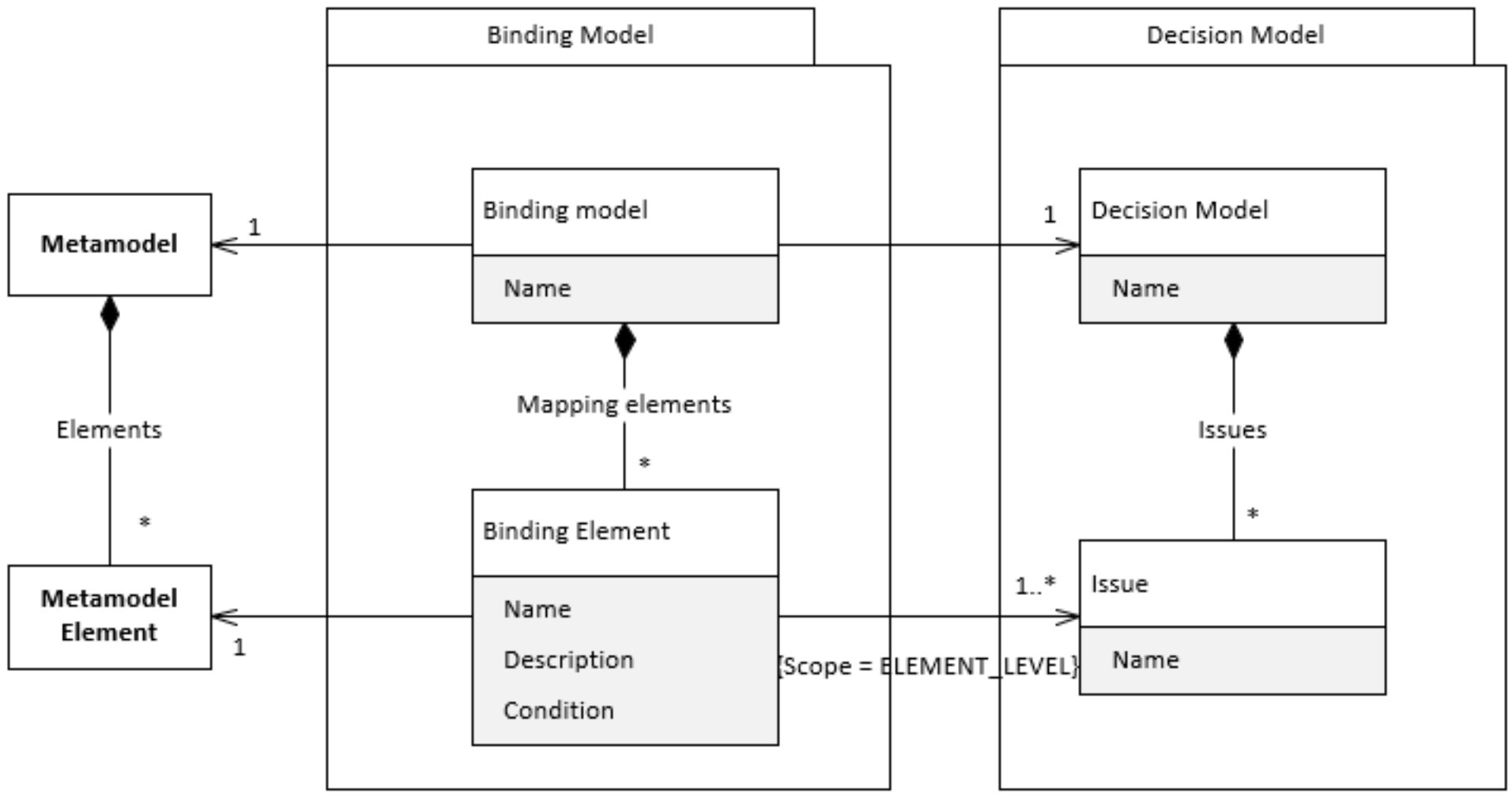

*Figure 4 Binding Model*

The root BindingModel class is a named container that references the source metamodel. The central class is BindingElement. Each BindingElement connects one or more design decisions to a metamodel concept and has a name, description, and an optional condition predicate specifying applicability criteria.

Each BindingElement has a sourceModelElementRef pointing to a metaclass (e.g., Entity or Relationship in an ER metamodel), realizing $\sigma_k \in MC$ from Definition 4. Bindings are defined only for source language concepts. Design decisions concern how to transform source concepts, regardless of the target platform.

Each BindingElement also has a decisionIssues collection referencing one or more Issue instances, realizing $I_k \subseteq I$ from Definition 4. A single BindingElement may reference multiple issues, and the same issue may appear in multiple BindingElements for different metamodel concepts.

The condition attribute realizes the predicate $\varphi_k$ as defined in Definition 4. When present, it is evaluated against source element properties at the M1 level, while the binding itself is defined at the M2 level in terms of metaclasses. When absent, the binding applies unconditionally.

The Binding Model distinguishes between element-level and model-level bindings. Element-level bindings reference ELEMENT_LEVEL issues, connect them to specific metaclasses, and may include conditions evaluated per source element. Model-level bindings reference MODEL_LEVEL issues that apply globally to the entire transformation. A model-level binding does not require a sourceModelElementRef or condition, as the associated decisions are resolved once for the entire source model rather than per element.

## 4.3. Configuration Model

The Configuration Model records the actual choices for a specific transformation task, realizing the configuration function (Definition 5) and validity conditions (Definition 6). Figure 5 presents the UML class diagram.

The root ConfigurationModel class has a sourceModel reference anchoring the configuration to a specific source model. The model has a two-tier structure mirroring the Decision Model's scope categorization.

GlobalChoice instances represent model-level decisions, each referencing an Issue and the selected Option (realizing $C_G : I_M \rightarrow O$ as defined in Definition 5). ElementChoice instances represent element-level decisions applied to specific source elements, each referencing the element, Issue, and selected Option (realizing the partial function $C_E : E \times I_E \rightharpoonup O$ from Definition 5 Both include optional attributes for documenting the reasoning behind selections (assumption, justification, consequences).

Different elements of the same type can receive different choices for the same issue. If the source model contains two 0..1 to 0..M relationships, the developer can independently choose ForeignKey for one and JoinTable for the other. This per-element configurability distinguishes our approach from global configuration switches.

In practice, the Configuration Model is generated automatically by analyzing the source model against the Binding Model, with defaults pre-selected per the Decision Model's isDefault flags. Proposition 2 guarantees the validity of the generated configuration. Developers then review and modify choices before execution.

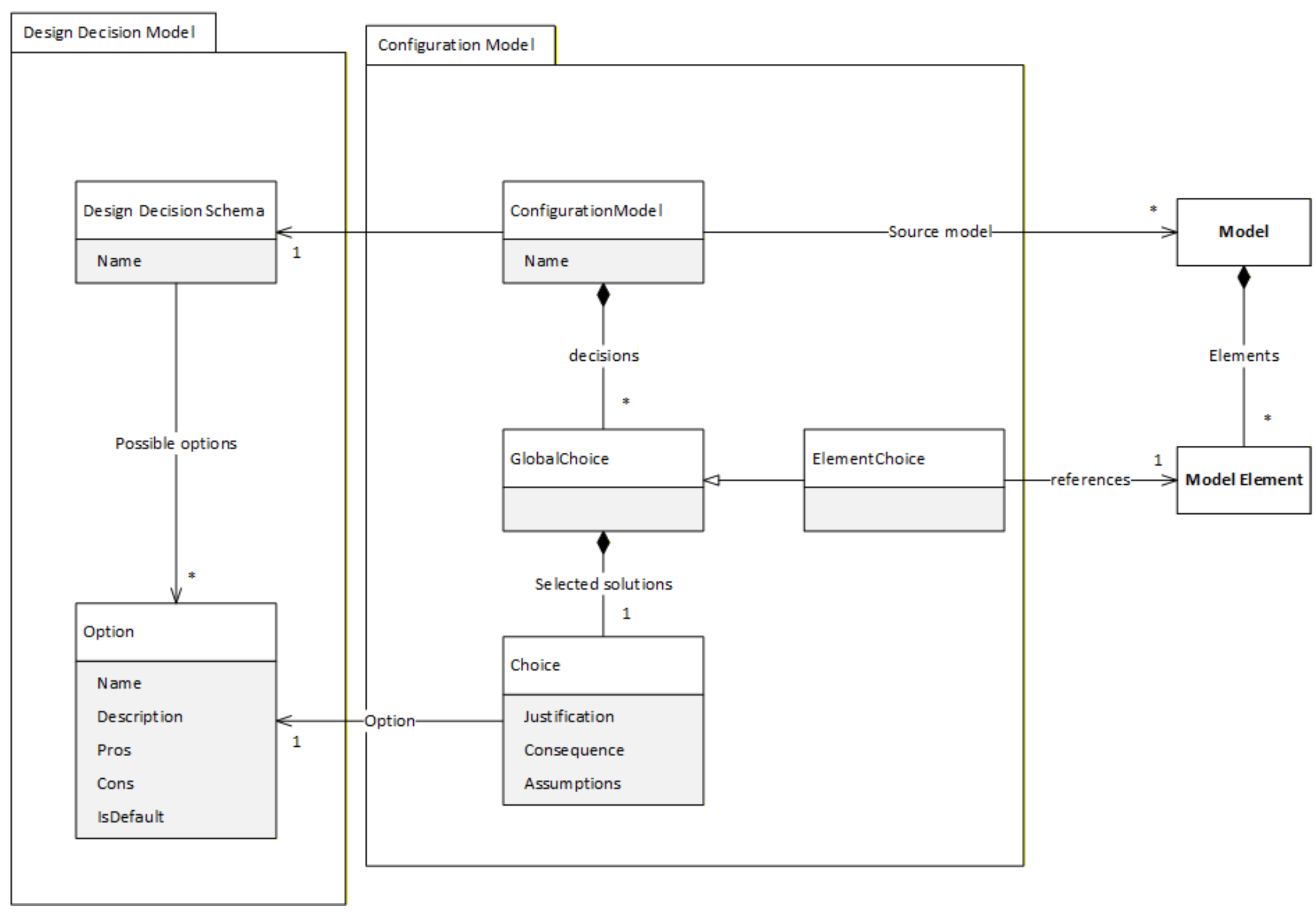

*Figure 5 Configuration Model*

## 4.4. Trace Model

The Trace Model is automatically generated during execution and implements the execution trace from Definition 12. It extends the formal structure with practical metadata and explicit references to the Configuration Model choices that guided rule selection. Figure 6 presents the UML class diagram.

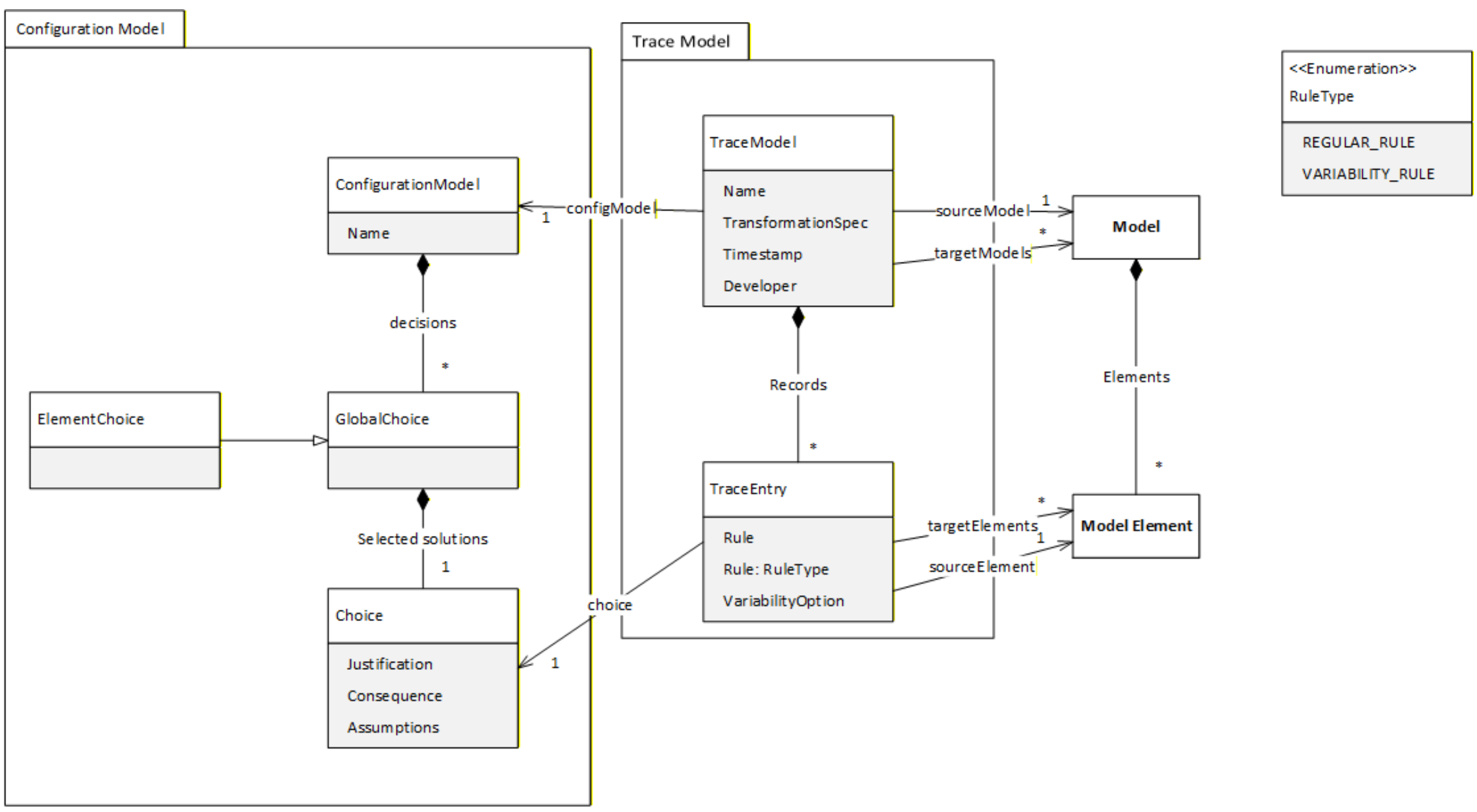

*Figure 6 Trace Model*

The root TraceModel class stores execution metadata (name, timestamp, developer, platformInfo) and maintains references to sourceModel, configModel, targetModel, and transformationSpec, ensuring that every execution can be traced and reproduced.

Each TraceEntry instance corresponds to one trace tuple. Each entry records the sourceElement, targetElements, rule applied, and ruleType (REGULAR_RULE or VARIABILITY_POINT per Definition 8). For variability point entries, variabilityOption records the selected variant (per the resolution function, Definition 11) and choice links to the Choice that guided resolution.

The Trace Model references transformation rules rather than containing copies of logic. The referencing mechanism is left to concrete implementations.

# 5. Realization in Synthesis

The formal framework of Section 3 is independent of any specific transformation language. In this section, we realize it concretely by extending Synthesis [18], an open-source embedded DSL for model-driven engineering in TypeScript, with support for design decisions and configurable variability in model-to-model transformations.

Synthesis is based on a functional approach to MDE formalized in a separate paper [3]. In that approach, models are not object graphs but values produced by evaluating model expressions, which are term-algebraic compositions of element creation constructors, reference retrievals, and embedded computations. JSX, the XML-like syntax extension to JavaScript, serves as the concrete syntax for these expressions. A JSX tag such as <Entity name="Person"> compiles to a function call that creates a typed model element, and nested tags express associations between elements. Embedded TypeScript expressions within curly braces provide computation operators for iteration, conditionals, and modular composition.

Model transformations in Synthesis are expressed as TypeScript classes. A transformation class is decorated with @M2M to declare a model-to-model specification, and its methods serve as transformation rules. Each rule is a model template, a function that takes a source element and returns a target model expression in JSX. The @E2E decorator registers a method as a traced element-to-element rule. Polymorphic dispatch over the source metamodel's subtype hierarchy is realized through @SpecPoint and @SpecOption decorators, which mark methods as dispatch targets and their type-specific specializations.

## 5.1. Base Synthesis Framework

We first summarize the base concepts needed to follow the code examples in Section 6, then show how the artifact models of Section 4 relate to them:

- **Metamodels as type schemas**. Metamodels are defined using type constructors for entity types, subtypes, arrays, optionals, and references.
- **Models as values**. Models conforming to these schemas are constructed by evaluating JSX expressions, where each tag creates a typed element and nesting expresses associations.
- **Templates and rules**. A model template is a TypeScript function whose body is a JSX expression with parameters. Transformation rules are templates whose parameter is a source model element.
- **Execution trace.** The framework automatically records source-to-target correspondences for every @E2E-decorated rule invocation.
- **Megamodels.** A megamodel is a model whose elements are themselves models, each potentially conforming to a different type schema.
- **Weaving models**. A weaving model is a model whose properties are typed references to elements in other models. Since models are themselves model elements in [3], these references can point to both models and elements within models through the same mechanism, with no restriction on the OMG MDA level of the referenced element

The artifact models of Section 4 fit naturally into these concepts. The Decision Model is a standalone model containing Issues and Options. The remaining three models are all weaving models:

- The Binding Model links elements across the source metamodel and the Decision Model. Its Binding elements hold references to metaclasses (M2 level) and to Issues from the Decision Model. Its condition predicates are evaluated on specific source model elements (M1 level). This cross-level weaving, connecting M2 and M1 within a single model, goes beyond the standard use of weaving models in the literature, where links typically connect elements at the same MDA level [40].
- The Configuration Model links M1 source model elements to M2-level Issues and Options from the Decision Model, recording the actual choices made for a specific transformation task.
- The Trace Model links M1 source elements to M1 target elements (standard same-level weaving) and additionally references configuration choices, closing the traceability chain across levels.

The overall four-model architecture of Figure 2 is itself a megamodel containing these interrelated sub-models. The cross-level nature of these weaving models is made possible by the uniform treatment of models and elements in [3], where a typed reference can point to any model element regardless of its MDA level. This means the variability extension described next builds on the mechanisms already defined in [3], without requiring new ones for model representation.

## 5.2. Variability Extension

The extension introduces two new decorators, @VarPoint and @VarOption, alongside modifications to the dispatch and trace mechanisms:

- **Variability point declaration.** A method decorated with @VarPoint declares a design decision point in the transformation, referencing an issue from the Decision Model. The method body carries no implementation. The engine resolves the variability at execution time by selecting one of the associated option rules.
- **Variability option rules**. Each @VarOption method provides one concrete transformation strategy for a variability point. It includes a predicate that references an option from the Decision Model, specifying under which configuration choice this strategy should be selected.
- **Predicate-driven dispatch.** When the transformation engine meets a variability point during execution, it retrieves the relevant choice from the Configuration Model and evaluates each option's predicate against it. The option whose predicate evaluates to true is selected and executed, realizing the resolution function (Definition 11). Theorem 1 guarantees that exactly one option is selected when the sufficient conditions are met.
- **Extended trace entries.** The Synthesis trace model is extended for variability point invocations to additionally record the resolved option and the specific configuration choice that guided selection.

# 6. Illustrative Example

In this section, we use the motivating example from Section 2, transforming the ER model describing *Person* and *Company* entities connected by an *Employment* relationship into a relational database schema. The description follows the lifecycle shown in Figure 2, progressing from source model definition through design time artifact creation to application time configuration, execution, and trace production.

The type schemas for the ER and Relational modeling languages, together with their corresponding JSX element type definitions, follow standard Synthesis patterns and are not reproduced here. They are available in the project repository [18] alongside the executable examples for all listings in this section.

## 6.1. Source Model

Listing 1 presents the source model as a JSX model expression that, when evaluated, produces the ER schema as a model value [3].

```
export const CompanySchema = () => {
  return (
    <EERSchema name="CompanySchema" MDA_level="M1">
      <Entity name="Person">
        <Attribute name="personId" isKey={true}>
          {/* Domain declarations omitted for brevity */}
        </Attribute>
        <Attribute name="name" isKey={false}>
          {/* Domain declarations omitted for brevity */}
        </Attribute>
      </Entity>
      <Entity name="Company">
        <Attribute name="companyId" isKey={true}/>
        <Attribute name="companyName" isKey={false}/>
      </Entity>
      <Relationship name="Employment">
        <Role
          name="WorksIn" lb="0" ub="1"
          domain={<Entity $refByName="Company"></Entity>}
        />
        <Role
          name="Hires" lb="0" ub="M"
          domain={<Entity $refByName="Person"></Entity>}
        />
      </Relationship>
    </EERSchema>
  )}
```

Listing 1. Source model: CompanySchema

The schema contains two entities, Person and Company, each with identifying and descriptive attributes. The Employment relationship connects these entities with a 0..1 to 0..M multiplicity pattern, meaning that each person works in at most one company and each company employs zero or more persons. Each Role uses $refByName to reference the participating entity by name, realizing the reference retrieval mechanism described in [3].

## 6.2. Design Time Artifacts

In the design time phase, three reusable artifacts are created. The Decision Model captures domain-level design decisions, the Binding Model connects those decisions to ER metamodel concepts, and the Transformation Specification implements the transformation rules with variability support.

### 6.2.1. Decision Model

Listing 2 presents the Decision Model for the relational schema transformation domain, realizing the design decision space from Definition 1.

```
export const RelationSchemaDesignIssues = (store: ModelStore) => (
  <DecisionModel
    name="RelationSchemaDesign"
    description="Design space for persistence strategies"
    store={store}
  >
    <ElementIssue
```

```
      name="OneToManyStrategy"
      description="How to implement 1:N relationships">
      <Option
        name="ForeignKey"
        description="Use Foreign Key in the Many side table"
        isDefault={true}
        optionType={<SimpleOption name="FK" />}
      />
      <Option
        name="JoinTable"
        description="Use a separate join table"
        optionType={<SimpleOption name="JoinTableOption" />}
      />
    </ElementIssue>
    <ElementIssue name="PrimaryKeyStrategy"
                  description="How to implement primary keys">
      <Option name="UseAutoIncrement"
              description="Database auto increment"
              isDefault={true}
              optionType={<SimpleOption name="PK_AutoInc" />}
      />
      <Option name="UseManualIncrement"
              description="Use manual increment"
              isDefault={false}
              optionType={<SimpleOption name="PK_ManInc" />}
      />
    </ElementIssue>
    <ModelIssue
      name="NamingConvention"
      description="How to name the database tables">
      <Option
        name="CamelCase"
        description="Use camelCase for table names"
        isDefault={true}
        optionType={<SimpleOption name="CamelCaseOption" />}
      />
      <Option
        name="SnakeCase"
         description="Use snake_case for table names"
        optionType={<SimpleOption name="SnakeCaseOption" />}
      />
    </ModelIssue>
  </DecisionModel>
)
```

Listing 2. Decision Model: RelationSchemaDesignIssues

The model defines three design issues. *OneToManyStrategy* addresses the central question from the motivating example, offering *ForeignKey* and *JoinTable* as alternatives for mapping 0..1 to 0..M relationships. *PrimaryKeyStrategy* provides two options for generating primary keys. *NamingConvention* offers two naming styles for generated database artifacts. Each issue declares exactly one default option via the *isDefault* attribute, satisfying the requirement from Definition 1 that $def(i) \in \delta(i)$ for all issues.

### 6.2.2. Binding Model

Listing 3 presents the Binding Model that connects the Decision Model from Listing 2 to the ER metamodel, realizing Definition 4.

```
export const ER_Bindings = (store: ModelStore) => (
  <BindingModel
    name="ER_Bindings"
    description="Binding EER Model concepts to relation schema design issues"
    sourceModel={<Model $refByName="/ERMetamodel" />}
    decisionModel={<DecisionModel $refByName="/RelSchemaDesign" />}
  >
    <Binding
      name="PKStrategyBinding"
      description="Binding Primary Key Strategy to EER Model"
      element={<Element $refByName="/ERMetamodel/Attribute" />}
      issues={[
        <ElementIssue $refByName="/RelSchemaDesign/PrimaryKeyStrategy" />,
      ]}
    />
    <Binding
      name="OneToManyBinding"
      description="Binding 1:N Relationship Strategy to EER Model"
      element={<Element $refByName="/CompanySchema/Relationship" />}
      issues={[
        <ElementIssue $refByName="/RelSchemaDesign/OneToManyStrategy" />,
      ]}
      condition='return (e) => {
                if (!e.roles || e.roles.length !== 2) return false;
                const m1 = e.roles[0];
                const m2 = e.roles[1];
                return (m1.ub === "1" && m1.lb === "0" && m2.ub === "M") ||
                       (m1.ub === "M" && m2.ub === "1" && m2.lb === "0");
            }'
    />
  </BindingModel>
)
```

Listing 3. Binding Model: ER_Bindings

Each <Binding> connects one or more design decisions to a metamodel concept. The element attribute identifies the metaclass (M2 level) to which the binding applies, and the issues attribute lists the relevant design decisions.

The first binding, *PKStrategyBinding*, connects the *Attribute* metaclass to *PrimaryKeyStrategy*. No condition is specified, so this binding applies unconditionally to all Attribute instances. The second binding, *OneToManyBinding*, connects the *Relationship* metaclass to *OneToManyStrategy* and includes a condition predicate that checks whether the relationship has exactly two roles with cardinalities matching the 0..1 to 0..M pattern. Only relationships that satisfy this condition will have the *OneToManyStrategy* decision applied during configuration generation.

### 6.2.3. Transformation Specification

The transformation specification implements the ER-to-Relational transformation as a TypeScript class. Listing 4 shows the class declaration.

```
@M2M()
class    EER2RelDomainWithDecisionTransformation    extends    abstractM2M
<IEERSchema, IRelSchema, {}, IConfigurationModel> {
    // transformation rules defined as methods
}
```
Listing 4. Transformation class declaration with @M2M decorator

The @M2M decorator declares that this transformation accepts source models conforming to the EER metamodel and produces target models conforming to the Relational metamodel. The generic type parameters bind the transformation to specific source and target metamodels (IEERSchema, IRelSchema). The fourth parameter, IConfigurationModel, is the extension introduced in this paper, making the configuration available to variability dispatch during execution.

Listing 5 presents the regular rule for transforming entities into tables.

```
@E2E()
Entity2Table(e: IEntity): Element<ITable> {
    return (
      <Table name={this.applyNaming(e.name)}>
          {e.attributes.map((a) => this.Attribute2Column(a))}
      </Table>
    )
  }
```
Listing 5. Regular rule: Entity to Table transformation

This rule creates a <Table> element for each Entity, applying the configured naming convention to the table name. It recursively invokes transformation rules for each attribute. The rule for transforming attributes into columns follows a similar pattern and is omitted for brevity.

The central design decision from the motivating example is realized as a variability point. Listing 6 shows the variability point rule, and Listing 7 presents its two variability option rules.

```
@E2E()
@VarPoint("OneToManyStrategy")
RelationshipMapping (rel: IRelationship): Element<ITable>|Element<IColumn> {
    // abstract, no implementation; resolved by the engine
}
```
Listing 6. Variability point rule for relationship mapping

The `@VarPoint` decorator marks this method as a variability point. As specified by the formal framework (Definition 7), variability point rules carry no implementation body. When the transformation engine encounters this variability point during execution, it evaluates the predicates of the associated variability option rules against the Configuration Model to determine which concrete strategy to apply.

```
@Default()
@VarOption("RelationshipMapping", Option("ForeignKey"))
relationshipAsForeignKey(rel: IRelationship): Element<IColumn> {
    const fkSide = this.getFKSide(rel)
    const pkSide = this.getPKSide(rel)
```

```
    return (
      <Table $refByName={this.applyNaming(pkSide.domain.name)}>
        <Column
          name={this.applyNaming(fkSide.domain.name + "Id_FK")}
          type={this.getColumnType(fkSide.domain)}
          isForeignKey={true}
          isKey={false}
          references={fkSide.domain.name}
          isNullable={fkSide.lb === "0"}
        />
      </Table>
    )
  }

@VarOption("RelationshipMapping",Option("JoinTable"))
relationshipAsJoinTable(rel: IRelationship): Element<ITable> {
    const role1 = rel.roles[0]
    const role2 = rel.roles[1]
    return (
      <Table name={this.applyNaming(rel.name)}>
        <Column
          name={this.applyNaming(role1.domain.name + "Id")}
          type={this.getColumnType(role1)}
          isForeignKey={true}
          references={role1.name}
          isNullable={role1.lb !== "1" && role1.ub !== "1"}
          isKey={role1.ub === "1"}
        />
        <Column
          name={this.applyNaming(role2.domain.name + "Id")}
          type={this.getColumnType(role2)}
          isForeignKey={true}
          references={role2.name}
          isNullable={role2.lb !== "1"}
          isKey={role2.ub === "1"}
        />
      </Table>
    )
  }
```

Listing 7. Variability option rules: foreign key and join table strategies

Each @VarOption decorator references the variability point by name and specifies a predicate that references an option from the Decision Model. The foreign key variant adds a nullable foreign key column to the table on the "many" side of the relationship. The join table variant generates a separate table with two foreign key columns referencing both participating entities.

Predicates can also be complex Boolean expressions combining multiple option references using Or(), And(), and Not(). A variability option annotated with @Default() serves as the fallback when no other predicate evaluates to true.

## 6.3. Application Time

The application time phase applies the design time artifacts to the CompanySchema source model. This phase involves configuration generation, transformation execution, and trace production.

### 6.3.1. Configuration Generation

Creating a Configuration Model manually for every source model would be tedious and error-prone, especially for large models with many elements. To address this, we implemented a dedicated transformation that takes the source model and the Binding Model as inputs and produces a default Configuration Model. The transformation analyzes each source element against the bindings, creates choices for all applicable design decisions, and pre-selects the default option for each. Proposition 2 guarantees that the resulting configuration is valid and can be used for execution without modification. The generation works in two steps.

First, the transformation identifies all model-level design decisions from the bindings. In this example, NamingConvention affects how all generated names are formatted globally, so a single global choice is created with the default option (CamelCase) pre-selected.

Second, the transformation goes through each element in the source model and evaluates the applicable bindings. For the Person entity, the PKStrategyBinding matches unconditionally, so a choice for PrimaryKeyStrategy is created with the default option (Auto Increment). The same occurs for the Company entity. For the Employment relationship, the condition predicate of OneToManyBinding is evaluated against the relationship's roles. Since the cardinalities match the 0..1 to 0..M pattern, the predicate evaluates to true, and a choice for OneToManyStrategy is created with the default option (ForeignKey). Listing 8 presents the resulting Configuration Model.

```
export const CompanySchemaConfig = (store: ModelStore) => (
    <ConfigurationModel
      name="CompanySchemaConfig"
      sourceModel={<Model $refByName="/CompanySchema" />}
    bindingModel={<BindingModel $refByName="/ER_Bindings" />}
    store={store}>    >
        {/* Global choices (model-level decisions) */}

<GlobalChoice name="CamelCaseNaming"
      choices={[
        <Choice name="UseCamelCase"
   issue={<GlobalIssue  $refByName="/RelationSchemaDesign/NamingConvention"
/>}
      selectedOption={<Option  $refByName="/RelationSchemaDesign/CamelCase"
/>}
        />,
      ]}
    /> ...

         {/* Element-specific choices */}
<ElementChoice name="PersonPrimaryKey"
      element={<Element $refByName="/CompanySchema/Person" />}
      choices={[
        <Choice name="UseAutoIncrement"
  issue={<ElementIssue
          $refByName="/RelationSchemaDesign /PrimaryKeyStrategy"/>}
  selectedOption={<Option  $refByName="/RelationSchemaDesign/AutoIncrement"
/>}
        />,
```

```
      ]}
    />

...
<ElementChoice name="EmploymentForeignKeyMapping"
      element={<Element $refByName="/CompanySchema/Employment" />}
      choices={[
        <Choice name="UseForeignKey"
          issue={<ElementIssue
                $refByName="/RelationSchemaDesign /OneToManyStrategy" />}
      selectedOption={<Option $refByName="/RelationSchemaDesign/ForeignKey"
/>}
        />,
      ]}
    />

</ConfigurationModel>
);
```

Listing 8. Auto-generated Configuration Model with default choices

The generated Configuration Model serves as a starting point for the application developer to review and modify. Suppose the developer decides that the Employment relationship should use a join table rather than a foreign key to avoid null values in the Person table. The developer also decides to use snake_case naming for compatibility with certain database conventions. Listing 9 shows the modified choices.

```
<GlobalChoice name="SnakeCaseNaming"
      choices={[
        <Choice name="UseSnakeCase"
            issue={<GlobalIssue
                      $refByName="/RelationSchemaDesign/NamingConvention"
/>}
           selectedOption={<Option
                  $refByName="/RelationSchemaDesign/SnakeCase" />}
        />,
      ]}
    />
...

<ElementChoice name="EmploymentJoinTableMapping"
      element={<Element $refByName="/CompanySchema/Employment" />}
      choices={[
        <Choice name="UseJoinTable"
          issue={<ElementIssue
               $refByName="/RelationSchemaDesign/OneToManyStrategy" />}
          selectedOption={<Option
                $refByName="/RelationSchemaDesign/JoinTable" />}
          assumptions="Join table avoids nulls…"
          justification="Join tables handle optional…"
          consequences="A new join table will…"
        />,
      ]}
    />
```

Listing 9. Developer-modified Configuration Model choices (excerpts)

The justification, assumptions, and consequences attributes in Listing 9 document the reasoning behind the choice.

### 6.3.2. Execution

Once the Configuration Model is finalized, the transformation engine executes the transformation specification against the *CompanySchema* source model. The engine navigates the source elements and invokes the appropriate rule for each. When the *Person* entity is met, the `entityToTable` regular rule (Listing 5) produces a `<Table>` element and recursively invokes rules for each attribute, producing corresponding `<Column>` elements. Since the configured *NamingConvention* is *SnakeCase*, the naming helper converts names accordingly (e.g., "personId" becomes "person_id").

When the Employment relationship is encountered, the engine reaches the RelationshipMapping variability point (Listing 6) and identifies its two option rules, relationshipAsForeignKey and relationshipAsJoinTable (Listing 7). It evaluates their predicates against the Configuration Model by looking up the choice for Employment on the OneToManyStrategy decision. Since the developer selected *JoinTable*, the predicate `Option("JoinTable")` evaluates to true, and the `relationshipAsJoinTable` variant is executed. Listing 10 presents the resulting target schema.

```
<RelationalSchema name="company_schema">
    <Table name="person">
        <Column name="person_id" isKey={true} type="INTEGER" />
        <Column name="name" type="VARCHAR(255)" />
    </Table>

    <Table name="company">
        <Column      name="company_id"      type="INTEGER"      isKey={true}
isAutoincrement={true} />
        <Column name="company_name" type="VARCHAR(255)" />
    </Table>

    <Table name="employment">
        <Column   name="person_id_fk"   type="INTEGER"   isForeignKey={true}
references="Person" isNullable={true} isKey={false}/>
        <Column   name="company_id_fk"   type="INTEGER"   isForeignKey={true}
references="Company" isNullable={false} isKey={true}/>
    </Table>
</RelationalSchema>
```

Listing 10. Target relational schema (join table strategy with snake_case naming)

Had the developer retained the default *ForeignKey* choice for *Employment*, no separate *Employment* table would be generated. Instead, a nullable *company_id_fk* foreign key column would be added to the *person* table. The same transformation code creates structurally different target models depending on the Configuration Model, without any modification to the transformation specification.

### 6.3.3. Trace Model

During execution, the transformation engine automatically creates the Trace Model. Listing 11 shows representative entries.

```
<TraceModel name="CompanySchema to CompanySchema_Relational_Schema_with_Decision_Model"
    timestamp="2025-06-01T12:00:00Z"
    source={<ERSchema $refByName="/CompanySchema" />}
    targets={[<RelSchema $refByName="/CompanyRelationalSchema" />]}
    config={<ConfigurationModel $refByName={"/CompanySchema_Config"} />}
    transformationSpec="CompanySchemaTransformation"
    store={store}>
 <TraceEntry
      source={<Entity $refByName="/CompanySchema/Person" />}
      targets={[<Table $refByName="/company_schema/person" />]}
      rule="Entity2Table"
      ruleType="RegularRule"
    />

<TraceEntry
      source={<Attribute $refByName="/CompanySchema/personId" />}
      targets={[<Column $refByName="/company_schema/person_id" />]}
      rule="Attribute2Column"
      ruleType="RegularRule"
    />
    {/* ... additional attribute trace entries ... */}


<TraceEntry
      source={<Relationship $refByName="/CompanySchema/Employment" />}
      targets={[<Table $refByName="/company_schema/employment" />]}
      rule="RelationshipMapping"
      ruleType="VariabilityPoint"
      variabilityOption="relationshipAsJoinTable"
      choices={[<Choice $refByName="/ConfigurationModel/UseJoinTable"
        issue={<ElementIssue
             $refByName="/RelationSchemaDesign/OneToManyStrategy" />}
        selectedOption={<Option
                       $refByName="/RelationSchemaDesign/JoinTable" />}
      />]}
    />
</TraceModel>
```

Listing 11. Trace Model (representative entries)

Each `<TraceEntry>` records the source element, the target element it was transformed into, the applied rule, and the rule classification. For regular rules such as `entityToTable`, the entry records the rule name and type. For variability points, the entry additionally records the resolved variability option and the Configuration Model choice that guided the selection.

The trace entry for the Employment relationship links the design decision (OneToManyStrategy in the Decision Model), through the developer's choice (JoinTable in the Configuration Model), to the applied rule variant (relationshipAsJoinTable), and to the produced target element (the employment table). The justification recorded in the Configuration Model is also accessible through this chain.

# 7. Related Work

## 7.1. Modeling Design Decisions in Software Development

The importance of design decisions in software development is well recognized, and many researchers advocate treating them as first-class artifacts [19], [20], [21], [22], [23], [24], [25]. When design decisions lack explicit representation, they become embedded in architecture and code, leading to "knowledge vaporization" [4]. Two main families of techniques address this problem. Variability modeling approaches, rooted in software product line engineering [7], [8], [27], represent decisions as variation points with selectable alternatives. Feature Models (FMs) [27] and their extensions [28] provide hierarchical specifications of system features, while Orthogonal Variability Models (OVMs) [30], [31] represent variability independently across multiple dimensions. A different line of work models design decisions explicitly as first-class artifacts with rationale documentation [25], [26]. Our Decision Model (Section 4.1) extends this idea by adapting it to the transformation domain, distinguishing between model-level and element-level decisions.

Comprehensive comparisons between feature modeling and explicit decision modeling are given in [32] and [33]. There have also been attempts to unify different variability models and demonstrate their equivalence with decision models [34], [35].

## 7.2. Mapping Design Decisions to Models

Könemann and Zimmermann [19] introduce a method for explicitly binding design decisions to model elements, ensuring that the rationale behind each design choice is accessible and reusable. However, this approach binds decisions directly to model elements at the M1 level. Our approach defines bindings at the M2 level (metamodel concepts), which are then resolved for specific model elements at the M1 level through the Configuration Model. This separation enables the reuse of bindings across different source models conforming to the same metamodel.

Czarnecki and Antkiewicz [36] focus on mapping FMs to other types of models, such as UML activity and class diagrams. This template-based approach superimposes all potential variants into a single model template, annotated with presence conditions and meta expressions using Boolean formulas over feature variables. Our predicate language for variability options (Definition 9) employs a similar mechanism of Boolean formulas, but applies them to guard transformation rule variants rather than model elements. Furthermore, this approach annotates variability inline within model templates, whereas our approach maintains design decisions as a separate artifact.

OVMs take a different approach than [36]. Rather than annotating model elements inline, OVMs maintain variability as a separate model that references the artifacts it affects. Our Binding Model (Section 4.2) shares this principle of maintaining the mapping as a separate artifact, but operates at the M2 level by binding design decisions to metamodel concepts rather than to specific model elements, and targets transformation rules rather than product configurations.

## 7.3. Supporting Design Decisions in Model Transformations

Two related lines of work address variability in transformations without explicitly supporting design decisions. Westfechtel and Greiner [37] extend single-variant model transformations to handle multi-variant models by reusing transformation traces to propagate annotations from source to target models. Strüber et al. [38] consolidate similar transformation rules into compact representations to improve maintainability, and extend algebraic graph transformations to handle variability in SPLE contexts. Both approaches improve variability handling within transformations. However, neither explicitly models the design decisions that determine how variability is applied, nor do they provide traceability from transformation outcomes back to the rationale behind specific mapping choices.

Sijtema [39] more directly supports design decisions within model transformations by extending ATL with variability rules that configure regular ATL rules, enabling them to be executed only when a corresponding feature is selected in the feature model. Higher-order transformations process variability-enriched ATL models into standard ATL models, ensuring backward compatibility with existing ATL engines. However, the approach lacks a built-in validation mechanism for checking annotations against feature constraints, and some aspects remain work in progress. Unlike our approach, this method couples variability to a feature model rather than to an independent decision model, does not distinguish between model-level and element-level decisions, and does not provide element-specific configuration.

None of the reviewed approaches treat design decisions as independent, first-class artifacts with formal properties, provide element-specific configuration, or offer a complete traceability chain from decisions through configuration to transformation outcomes. Furthermore, our artifact models (Binding, Configuration, Trace) are realized as weaving models in the sense of [3], linking elements across different models and MDA levels. This cross-level weaving goes beyond the standard use of weaving models in the literature [40], where links typically connect elements at the same abstraction level.

# 8. Conclusion and Future Work

MDE promotes the idea that models should be the central artifacts of software engineering. Yet, model transformations, the mechanisms that give models their operational power, typically do not make explicit the design decisions they embody. This paper has presented an approach to addressing this limitation by treating design decisions as first-class artifacts in rule-based model transformations.

## 8.1. Contributions

The core problem is the embedding of design decisions within transformation code, which leads to rigid, difficult-to-maintain, and poorly documented transformations. We have made four contributions:

(1) A formal mathematical framework (Section 3) that defines the core concepts of variability-based transformations through 12 definitions, 2 theorems, 2 propositions, and a corollary, proving key properties including well-definedness, separation, and conservative extension.
(2) A practical architecture of four interconnected artifact models (Section 4) that realizes the formal framework, with the Binding, Configuration, and Trace models realized as cross-level weaving models.
(3) A validation through an implementation that extends Synthesis with variability support (Section 5).
(4) A complete ER-to-Relational illustrative example (Section 6) covering the full lifecycle from source model definition through configuration to execution and trace generation.

## 8.2. Limitations

The most significant limitation is the upfront modeling effort required to define the Decision Model, create Binding Models, and implement transformation rules with variability points and option structures. This investment may not be justified for simple, one-time transformations, but it becomes increasingly valuable as transformations are reused or maintained over extended periods.

There is also a learning curve associated with adopting the approach. Transformation engineers must understand the formal framework, the four-model architecture, and their relationships. Comprehensive documentation, worked examples, and tooling support can mitigate this challenge.

The current implementation is specific to AelasticS Synthesis. While the formal framework and model architecture are applicable to other transformation environments, adapting them to languages such as ATL or QVT would require significant effort. Languages supporting decorators (Python) or annotations (Java, C#) could adopt the approach more readily.

For large source models with many element-level decisions, manual review of the generated configuration can become burdensome, even with defaults pre-selected.

## 8.3. Future Work

Several directions for future research emerge from this work.

The Decision Model currently treats design issues and their options independently, with no way to express dependencies between them. In practice, selecting certain options may exclude or require others. Extending the model with formal constraint specifications would enable automatic consistency checking during configuration creation, preventing invalid combinations.

Another approach is to use generative AI to reduce the manual effort required to create artifact models. Generative AI could analyze existing transformation code to extract implicit design decisions and generate initial Decision Model structures, propose Binding Models for new source languages through metamodel analysis, and suggest Configuration Model contents based on domain knowledge. In all cases, developers would review and adjust the generated artifacts.

Extending the approach to bidirectional and incremental transformations presents further challenges. Investigating how design decisions propagate in both directions and how configurations should be updated when models change incrementally would broaden the applicability of the approach.

Finally, empirical evaluation with industrial case studies would provide valuable evidence regarding practical benefits and adoption challenges.

In conclusion, this paper demonstrates that design decisions in model transformations can be explicitly modeled and effectively managed, enabling more flexible, traceable, and maintainable transformation systems.